\documentclass[utf8]{frontiersFPHY} 

\setcitestyle{square} 
\usepackage{url,hyperref,lineno,microtype,subcaption}
\usepackage[onehalfspacing]{setspace}
\usepackage{aas_macros}

\newcommand{\numax}{\nu_{\mathrm{max}}}
\newcommand{\teff}{T_{\rm eff}}
\newcommand{\yini}{Y_{\mathrm{ini}}}

\newcommand{\fov}{f_{\mathrm{ov}}}
\newcommand{\tauheii}{\tau_{\mathrm{HeII}}}
\newcommand{\tdyn}{t_{\mathrm{dyn}}}
\newcommand{\chifreq}{\chi^2_{\mathrm{freq}}}
\newcommand{\chispec}{\chi^2_{\mathrm{spec}}}
\newcommand{\chidyn}{\chi^2_{\mathrm{dyn}}}
\newcommand{\Kepler} {\textit{Kepler}\;}

\def\keyFont{\fontsize{8}{11}\helveticabold }
\def\firstAuthorLast{P\'erez Hern\'andez {et~al.}} 
\def\Authors{F. P\'erez Hern\'andez\,$^{1,2,*}$, R. A. Garc\'{\i}a\,$^{3,4}$, S. Mathur\,$^{1,2}$, A. R. G. Santos\,$^{5}$, 
and C. R\'egulo\,$^{1,2}$}
\def\Address{$^{1}$Instituto de Astrof\'{\i}sica de Canarias, La Laguna, Tenerife, Spain\\
$^{2}$Dpto. de Astrof\'{\i}sica, Universidad de La Laguna, La Laguna, Tenerife, Spain \\
$^{3}$IRFU, CEA, Universit\'e Paris-Saclay, Gif-sur-Yvette, France\\
$^{4}$AIM, CEA, CNRS, Universit\'e Paris-Saclay, Universit\'e Paris Diderot, Sorbonne Paris Cit\'e, Gif-sur-Yvette, France\\
$^{5}$Space Science Institute, 4750 Walnut Street, Suite 205, Boulder CO 80301, USA}
\def\corrAuthor{Instituto de Astrof\'{\i}sica de Canarias, E-38205 La Laguna, Tenerife, Spain}

\def\corrEmail{fph@iac.es}

\begin{document}
\onecolumn
\firstpage{1}

\title[Magnetic activity and asteroseismic determinations]{Influence of magnetic activity on the determination of
stellar parameters through asteroseismology}

\author[\firstAuthorLast ]{\Authors} 
\address{} 
\correspondance{} 

\extraAuth{}

\maketitle

\begin{abstract}

Magnetic activity changes the gravito-acoustic modes of solar-like stars and in particular their frequencies. There is an angular-degree dependence that is believed to be caused by the non-spherical nature of the magnetic activity 
in the stellar convective envelope. These changes in the mode frequencies could modify the small separation of low-degree modes (i.e. frequency difference between consecutive quadrupole and radial modes), which is sensitive to the core structure and hence to the evolutionary stage of the star. Determining global stellar parameters such as the age using mode frequencies at a given moment of the magnetic activity cycle could lead to biased results. 
Our estimations show that
in general these errors are lower than other systematic uncertainties, but in some circumstances they can be as high
as $10\%$ in age and of a few percent in mass and radius. 
In addition, the frequency shifts caused by the magnetic 
activity are also frequency dependent. In the solar case this is a smooth function that will mostly be masked by the filtering of the 
so-called surface effects. However the observations of other stars suggest that there is an oscillatory component with a period close to the one corresponding to the acoustic depth of the  He$\,$II zone. This could give rise to a misdetermination of some global stellar parameters, such as the helium abundance. Our computations show that the uncertainties introduced by this effect are lower than the $3\%$ level.

\tiny
 \keyFont{ \section{Keywords:} stars , asteroseismology , stellar magnetic activity , KIC~8006161, KIC~9139163}
\end{abstract}

\section{Introduction}

Stars located below the instability strip in the Hertzsprung-Russell diagram have external convective zones and are usually 
called solar-like stars. Turbulent motions inside these convective layers excite gravito-acoustic modes \citep{1977ApJ...212..243G} that are used in asteroseismology to characterize the internal structure and dynamics of 
the stars. With the development of continuous (from months to years) high-precision photometry from space with several missions 
such as CoRoT \citep{2006cosp...36.3749B}, \Kepler \citep{2010Sci...327..977B}, K2 \citep{2014PASP..126..398H}, and TESS \citep{2014SPIE.9143E..20R} hundreds of main-sequence stars and dozens of thousands red giants have been observed.  

For many of these solar-like stars, the fundamental parameters have been estimated using the seismic measurements. Using the 
so-called ``global asteroseismic scaling relations'' \citep[e.g.][]{1991ApJ...368..599B,1995A&A...293...87K,2011ApJ...732...54C}, masses and radii are obtained with a typical precision of 
10\% and 5\% respectively \citep[e.g.][]{2010A&A...509A..77K}. These estimates are model-independent because they rely on three 
observables: the effective temperature of the star, $\teff$, the frequency of the maximum power where the modes are located, 
$\numax$, and the large frequency spacing, $\Delta\nu$. The observed data from the Sun and other stars are used as references.
Better results can be obtained using stellar models to derive mass, 
radius, and also age. In the case of grid-based modeling, when fitting spectroscopic observables and global seismic parameters 
all together to a pre-calculated grid of models, we can reach a precision better than 3\% in radius, 6\% in mass, and 23\% in 
age \citep{2017ApJS..233...23S}. The improvement is even better when incorporating the information from the individual mode 
frequencies with systematic uncertainties of 1\% in radius, 3\% in mass, and 15\% in age \citep[e.g.][]{2012ApJ...749..152M,2014EAS....65...99L,2014EAS....65..177L,2014ApJS..214...27M,2015MNRAS.452.2127S,2017A&A...601A..67C}.

The precision reached by the asteroseismic inferences when the information of the individual modes is used, is among the best 
for field stars. Therefore, they are particularly interesting for other research fields such as galactic-archaeology studies 
\citep[e.g.][]{2013MNRAS.429..423M,2015ASSP...39...11M,2016A&A...597A..30A} as well as to properly characterize the exoplanet radius through the transit method \citep[e.g.][]{2013ApJ...767..127H,2018MNRAS.tmp.1712V,2018MNRAS.478.4866V}. 

The combined action of convection and differential rotation in main-sequence solar-like stars can produce magnetic fields of 
dynamo origin \citep[e.g.][]{2017LRSP...14....4B}. Under the action of stellar winds, stars of masses below 
$\sim 1.4 M_{\odot}$ slow down during their evolution
\citep[e.g.][]{1972ApJ...171..565S,1988ApJ...333..236K,2012ApJ...754L..26M,2014A&A...572A..34G,2015ApJ...798..116R}. Sometimes, these magnetic fields change regularly, producing magnetic cycles similar to the eleven-year cycle of the Sun in which the poles change polarity, requiring 22 years to complete a full magnetic cycle. 
The perturbations induced by the magnetic fields modify the properties of the acoustic modes measured in the Sun \citep[e.g.][]{1985Natur.318..449W,1989A&A...224..253P,2014SSRv..186..191B} and in 
other stars \citep[e.g.][]{2010Sci...329.1032G,2016A&A...589A.118S,2017A&A...598A..77K,2018ApJS..237...17S,2018ApJ...852...46K}. In particular, mode frequencies are shifted towards higher frequencies as the magnetic cycle evolves towards 
the maximum activity. 

The shifts in the mode frequencies can introduce systematic errors in the inferred global stellar parameters, specially 
if the star is only observed during the maximum of an on-going magnetic activity cycle. However, 
if the frequency shifts induced by the magnetic field smoothly increase with the frequency, as it is in fact the dominant 
change in the Sun, this effect will be partially or perhaps mostly masked by the filtering of the so-called surface effects
(see \citep{2017MNRAS.464.4777H} where the time variation in the surface term due to the solar activity cycle was analyzed 
in detail).

Such filtering is required at present in any model fitting procedure to deal with the unknown physics of the upper layers
of the star \citep[e.g.][]{2008ApJ...683L.175K,2017A&A...600A.128B}. This is perhaps one of the reasons why these potential systematic uncertainties have not drawn great attention to the 
asteroseismic community.

However, there are additional terms in the frequency shifts induced by the magnetic activity, already present in the Sun and 
that seem to be more important in other stars. 
As it is going to be described below, the magnitude of the shifts in some stars not only are much larger than in 
the Sun but the shifts of the different low-degree modes are clearly different. This can introduce a bias in the stellar 
parameter determination.
Moreover, as uncovered by \cite{2018A&A...611A..84S}, in some stars the variation of the frequency shifts with frequency 
does not follow a simple smooth function but has a more sinusoidal shape, suggesting that the magnetic perturbation is occurring 
in deeper layers compared to the Sun. This frequency changes are 
 not filtered with the surface correction terms.

In this paper we study the impact of both effects in the extraction of the global stellar parameters providing additional 
uncertainties that takes into account these biases induced by the magnetic field on the frequencies. The observed and simulated 
data used are described in Section~2 whereas in Section~3 we explain in detail the frequency 
shifts considered in this work. In Section~4 we describe the methodology used to perform the model fitting. 
The results are discussed in Section~5 and some conclusions are presented in Section~6.

\section{Observed and simulated data} 

We have selected two main-sequence stars studied by \cite{2018A&A...611A..84S} and that show frequency changes due to magnetic activity. The first one is KIC$\,$8006161 that has about 
$1M_{\odot}$ and a radiative core. The second one is KIC$\,$9139163 which is a more massive star, of about $1.4M_{\odot}$, 
and presumably with a convective core.
The former star seems an ideal choice for a solar analog because it has a large range of detected modes, including some $\ell=3$, 
and it shows frequency changes correlated with the photometric magnetic activity 
that depends on angular degree $\ell$ \cite{2018ApJS..237...17S} and frequency $\nu$ \cite{2018A&A...611A..84S}.
The frequency range of the acoustic modes detected 
in KIC$\,$9139163 is smaller, but it corresponds to the typical number of modes we have for an F8 main-sequence star 
without mixed modes. This star is still perfect for our analysis as it shows both angular degree 
and frequency dependent shifts (see \cite{2018ApJS..237...17S} and \cite{2018A&A...611A..84S} respectively).
The spectroscopic parameters and frequency sets used for both stars are given in Table~\ref{table:1}. 
For KIC$\,$8006161 the full set of frequencies reported by 
\cite{2012A&A...543A..54A} includes $\ell=0$ modes with radial orders in the range $n=15-30$, $\ell=1$ modes with
$n=13-29$, $\ell=2$ modes with $n=15-29$ and $\ell=3$ modes with $n=19-24$. 
For KIC$\,$9139163 the range of frequencies considered was: $\ell=0,\, n=11-29$, $\ell=1,\, n=11-29$ and $\ell=2,\, n=10-28\,$.
The values of the frequency of maximum power $\numax$ and the large separation $\Delta \nu$ were not used in the fitting 
procedure but are indicated in Table~\ref{table:1} for guidance. 

The effect or bias that magnetic activity can introduce on the determination of the fundamental stellar parameters 
will be influenced by the observational constraints (spectroscopic and asteroseismic inputs) and the
stellar evolution and oscillation codes considered (underlying physics, free parameters considered, etc.). 
Since observations and theoretical models can be improved in the near future
we have found convenient to partially remove such effects by considering in addition to the real stars, 
two models as proxy stars. 
They are constructed from the best-fit model for each star (obtained as indicated in Section ~4). We used the output of the 
best-fit model 
as input parameter for computing the frequencies of the corresponding proxy star. In the minimization procedure the same 
set of $(n,\ell)$ modes were considered.
Although these proxies will be free of systematic errors other than the bias introduced by the frequency shifts from the 
simulated magnetic activity (to be discussed in the next section), 
assuming these targets free of observed errors would be unrealistic. So we have adopted the criteria described below.

\begin{figure}[h!]
\begin{center}
\includegraphics[scale=0.5]{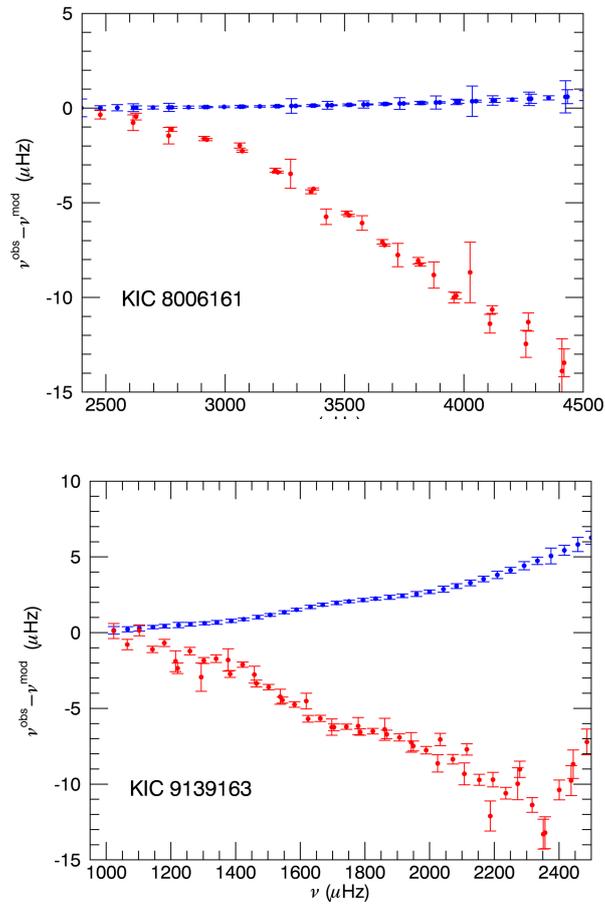}
\end{center}
\caption{Frequency differences between the observations (resp. proxy stars) and the best-fit model are represented with the red symbols (resp. blue symbols). The differences for the proxy stars are due 
 to the different boundary conditions used, as detailed in Section 2.}
\label{fig:1}
\end{figure}

Considering the theoretical frequencies for the 
proxies would assume that near-future models will introduce systematic errors much lower than the observed ones. 
Presumably this would not be the case
due to the complex physics required to model the upper layers both in the stellar structure models and in 
the oscillation laws for sonic, turbulent, non-adiabatic motions.
Although some progress has already been done for improving the physics of the uppermost layers (e.g. \cite{2017MNRAS.466L..43T}),
we can only expect that in the coming years surface effects would introduce diminished discrepancies between theoretical
and observational frequencies,
but remaining higher than the observational errors. Hence it is convenient that our proxy stars include some departure from the
models which gives rise to similar surface effects.
In particular we computed the eigenfrequencies of the proxy stars with a different surface 
boundary condition, namely $\delta P=0$ rather than the standard match to the analytical solution for an isothermal atmosphere at the top
of the model. 
Blue points in Figure~\ref{fig:1} correspond to the differences between the proxy stars (the same models with $\delta P=0$ as boundary condition) and the model frequencies (computed with the standard 
boundary conditions). 
Upper panel is for KIC$\,$8006161 and bottom panel for KIC$\,$9139163.
As expected these differences show a typical smooth frequency function with $\delta \omega \to 0$ as $\omega \to 0$ as any
other surface effect. In the same figure we have compared such effect with 
the surface terms found for the actual stars after the standard model fit procedure detailed in the Section~4 was carried 
out. The red points correspond to the differences between the observed frequencies and the ones from the best
models. Note that in our simulation the surface 
errors are of opposite sign but we have not found any indication of this being relevant in the final results.

Regarding the frequency errors in the proxies 
we have considered two cases: one case with the observed errors and the other one with half of the errors.
Similarly, theoretical spectroscopic parameters were used but with errors half the observed ones for each 
star; these inputs are indicated in Table~\ref{table:1}. 
The error in the frequencies are given by a combination of the frequency resolution, the signal-to-noise ratio of the 
spectrum, and the width of the modes 
(see  e.g. \cite{1992ApJ...387..712L}, \cite{1994A&A...289..649T}).
Maintaining the width and the SNR constant the error will be reduced as $\sqrt{T}$ where $T$ is the length of the observations. 
Thus, to reduce the error by a factor of 2, it is necessary to measure 4 times longer.

\section{Frequency shifts induced by the magnetic activity}

\subsection{Range of frequency shifts observed in main-sequence stars from \Kepler}

In \cite{2018ApJS..237...17S}, temporal variations of the mode frequencies were measured for 87 solar-like stars observed by 
\Kepler. Significant frequency shifts were found for more than half of the stars. Within the sample,
the observed absolute frequency shifts vary between $0$ and $2.2\,\mu\text{Hz}$ with a median value of $0.15\,\mu\text{Hz}$. 

\begin{figure}[h!]
\begin{center}
\includegraphics[scale=0.75]{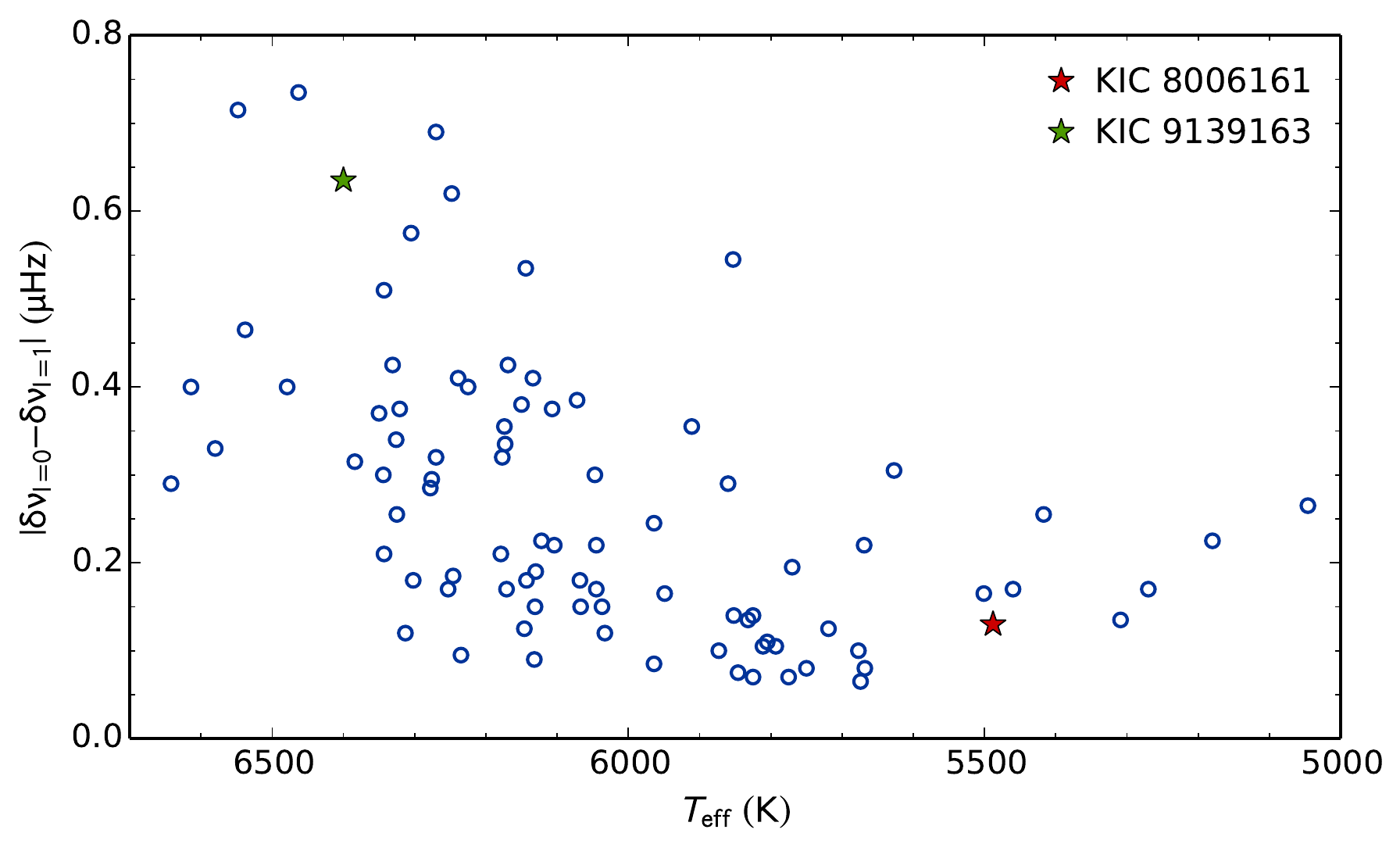}
\end{center}
\caption{Median absolute difference between the frequency shifts for $l=0$ and $l=1$ modes, 
$|\delta\nu_{l=0}-\delta\nu_{l=1}|$, as function of the effective temperature, $\teff$ for the 87 solar-like stars in 
\cite{2018ApJS..237...17S}. The red and green stars mark KIC$\,$8006161 and KIC$\,$9139163 respectively.}
\label{fig:2}
\end{figure}

As modes of different angular degree are sensitive to different latitudes and, in principle, magnetic activity is not uniformly 
distributed over the stellar surface, the frequency shifts for the radial and non-radial modes are not expected to be the same. 
In fact based on the results of \cite{2018ApJS..237...17S}, the difference between the frequency shifts for $l=0$ and $l=1$ 
modes is found to increase 
with effective temperature. However, we note that the mode frequencies are less accurately determined for hotter stars 
than cooler stars. 
Also, generally, the number of visible modes decreases with effective temperature. Figure~\ref{fig:2} shows the median 
difference in the frequency shifts as a function of the effective temperature for the 87 \Kepler targets. The stars 
KIC$\,$8006161 and KIC$\,$9139163 are highlighted as they are the focus of this study.

\begin{figure}[h!]
\begin{center}
\includegraphics[scale=0.5]{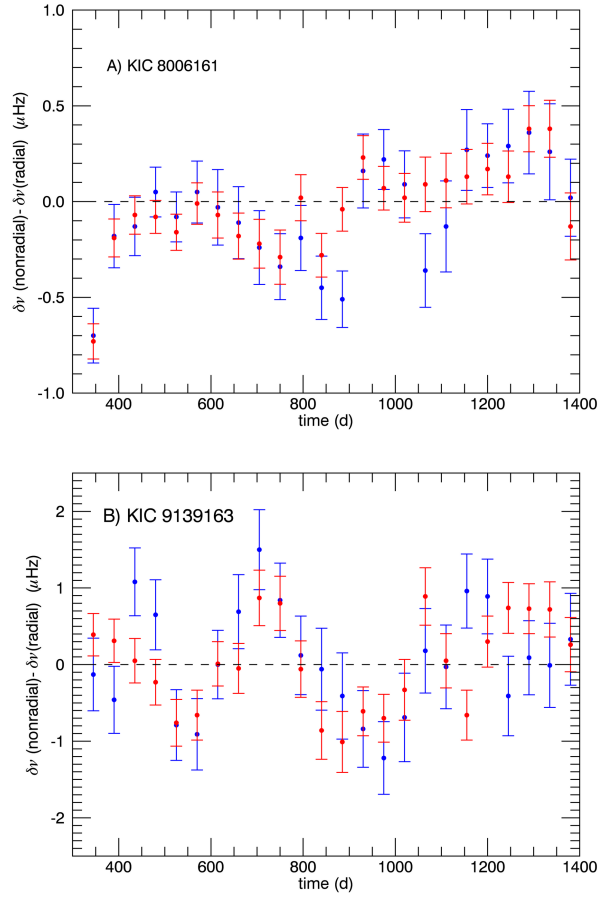}
\end{center}
\caption{Frequency shift differences $\delta\nu (\ell=1)-\delta\nu (\ell = 0)$ (red points) and
$\delta\nu (\ell=2)-\delta\nu (\ell = 0)$ (blue points) as a function of time (in days) for KIC$\,$8006161 (upper panel) and 
KIC$\,$9139163 (lower panel).}
\label{fig:3}
\end{figure}

Figure~\ref{fig:3} shows frequency shift differences between radial and non-radial modes as a function of time for the two stars 
considered in this work. Data were taken from \cite{2018ApJS..237...17S}. 
The red points correspond to the difference between the frequency shifts of $l=0$ and $l=1$ modes whereas the blue points
are the difference between the $l=0$ and $l=2$ modes. Although not shown in the figure, we note that 
for KIC$\,$8006161 the mean frequency shift of the radial oscillations remains almost constant in the
first half of the observed period (for the first 800 days) and then monotonically increases with time (where magnetic activity is increasing). From Figure~\ref{fig:3}
it seems that at this later stage of the activity cycle the shifts of the non-radial modes are systematically higher than the 
radial ones by tenths of $\mu$Hz whereas in the first
period the shift of the non-radial modes is similar or lower than for the radial ones.
As shown in Figure~\ref{fig:2} the range in the median differences between the shifts of the $\l=0$ and $\l=1$ modes for 
KIC$\,$9139163 is higher, but from Figure~\ref{fig:3} we can see that the periods when the non-radial shifts remain 
systematically higher or systematically lower than the radial ones are shorter. As a conclusion we can expect that frequency 
modes determined from observed periods of few hundreds of days can introduce differences in the frequency shifts between radial 
and non-radial modes of a few tenths of $\mu$Hz.

\subsection{Frequency shifts implemented in this work}

We have considered two different kinds of frequency shifts that can bias the determination of the stellar parameters 
based on standard stellar codes.
The first one is a frequency-dependent shift. As indicated in the introduction this kind of shift 
was observed in the Sun long time ago but has also been detected in other stars. In the Sun, the frequency dependence of the 
normalized frequency shifts $I_{nl} \delta \nu_{nl}$ where
$I_{nl}$ is the mode inertia of the mode with radial order $n$ and degree $l$ 
(see \cite{2010aste.book.....A} for a definition of mode inertia),
has a dominant smooth term, and hence its effects on the determination of the model parameters will be mostly masked 
by the filtering of the unknown surface term. However, as commented before, following the work by \cite{2018A&A...611A..84S},
in other stars there are some indications that 
the frequency dependence could include an oscillatory term with a period that could correspond to the acoustical depth of the 
He II zone. This term in general is not filtered out in the model fitting procedure.

\begin{figure}[h!]
\begin{center}
\includegraphics[scale=0.5]{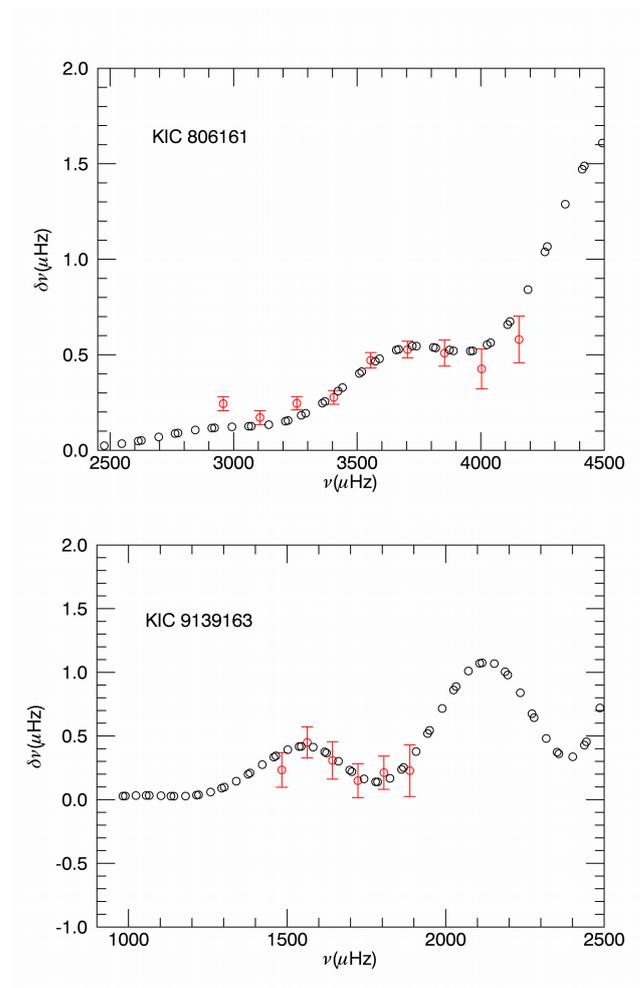}
\end{center}
\caption{Frequency-dependent shifts introduced in the simulations for both stars. 
Red symbols are the observed frequency differences computed as reported in \cite{2018A&A...611A..84S} for KIC$\,$8006161
whereas black symbols correspond to the full observed  mode set used in the modelling.}
\label{fig:4}
\end{figure}

To introduce these frequency shifts in our simulation we follow \cite{2016A&A...589A.118S} and \cite{2018A&A...611A..84S}
and consider a fit to the observed normalized frequency shifts of the form
\begin{equation}
I_{nl} \frac{\delta \nu_{nl}}{\nu_{nl}} \to 
A_0 + A_1 \cos \big(2\omega \tau + \varphi \big)
\; ,
\label{Eq:1}
\end{equation}
where $\omega= 2\nu$ is the angular frequency and $\tau$ the acoustical depth representative of the perturbation induced by 
the magnetic activity. Mode inertias $I_{nl}$ are interpolated from the best-fit models.
In principle there are a total of four parameters to be fitted $A_0$, $A_1$, $\tau$ and $\varphi$.
However, to minimize the errors, the mean values of the frequency shifts between adjacent $l=0$ and $l=1$ modes were considered
in \cite{2018A&A...611A..84S} and here we proceed in the same way. As a consequence of 
the small number of frequencies available, determining  $\tau$ strongly depends on the initial values considered. 
Therefore the parameter $\tau$ was fixed to $\tauheii$, the acoustic depth of the He~\textsc{ii} 
ionization zone given in Table~\ref{table:2} for each star, and, hence, a simple
linear fit with the parameters $A_0$, $A_1$ and $\varphi$ was considered.
The normalized frequency shifts thus corrected by the mode inertia were shown in Figure~10 of \cite{2018A&A...611A..84S} for
KIC$\,$8006161 and other stars.
We show in Fig.~\ref{fig:4} the absolute frequency shifts, $\delta \nu$, for our target stars (red points with yerror bars). 
The range of frequencies for which frequency 
shifts could be determined in \cite{2018A&A...611A..84S} is much smaller than the full set of frequencies used in the 
modelling and indicated above. Hence, we extrapolate the function given by Eq.~\ref{Eq:1}
to the full range of observed frequencies, resulting in the shifts represented by the black points in Fig.~\ref{fig:4}.

We should be aware that the frequency shifts derived from the \Kepler mission correspond to differences between the maximum and 
minimum of the magnetic activity observed in the \Kepler data, hence,they
do not necessarily correspond to differences between maximum and minimum of the stellar activity cycle.
In order to consider different situations, we have taken
changes in the amplitude of the oscillation component from $-5A_1$ to $5A_1$, $A_1$ being the value derived from the fit for
each star as given in Eq.~\ref{Eq:1}. In Section~5, where we show the results, 
we will see that the highest amplitudes turn out to be very large, but they cannot be rejected a priori.

The second kind of frequency shifts considered is an $\ell$-dependent frequency shift consistent with the observed values 
shown above. Specifically, in our simulation we have assumed that the observations are done at a given moment of the magnetic 
activity cycle with a time duration of at least several months. Here we add an $\ell$-dependent frequency shift
$\delta \nu_{\ell}$ to the term $\delta \nu_{n\ell}=A_0\nu_{nl}/I_{nl}$ defined in Eq.~\ref{Eq:1} and representing an overall
frequency increase with magnetic activity. 
We consider different cases, each one with a constant shift of $\delta \nu =0.2 f\, \mu$Hz for 
the $\ell=0$ and $\delta \nu = -0.1f\,\mu$Hz for
the non-radial oscillations, where $f$ is a free parameter that goes from $-1.5$ to $1.5$. This means that in the simulation the difference between the radial and non-radial frequency shifts can be as large as $\pm 0.45\mu$Hz. 
As noted above for a hot star like KIC$\,$9139163 the shifts can be even higher (larger than $1\mu$Hz) but such 
differences should be partially averaged out if periods longer than about 100 days are considered.

\section{Fitting procedure}

Model fitting is based on a grid of stellar models evolved from the pre-main sequence to the TAMS using 
the MESA code \cite{2011ApJS..192....3P}, \cite{2013ApJS..208....4P}, \cite{2015ApJS..220...15P}, version 10$\,$398.
The OPAL opacities \cite{1996ApJ...464..943I}, the GS98 metallicity mixture \cite{1998SSRv...85..161G} and 
the exponential prescription for the overshooting given by \cite{2000A&A...360..952H} were used, 
otherwise the standard input physics from MESA was applied. 
Eigenfrequencies were computed in the adiabatic approximation using the ADIPLS code \cite{2008Ap&SS.316..113C}.

As specifically implemented in MESA, the overshooting formulation corresponds to that introduced by 
\cite{2000A&A...360..952H} and it is a simplified version of the formulation given by \cite{1996A&A...313..497F}.
Here the overshooting parameter $\fov$ is defined by the equation 
\begin{equation}
D_{\mathrm{ov}} = D_0 \exp \left(\frac{-2z}{H_v} \right) \quad ; \quad H_{v}=f_{\mathrm{ov}} H_p
\; ,
\label{eq_over}
\end{equation}
where $D_\mathrm{ov}$ is the diffusion coefficient, $D_0$ the diffusion coefficient 
at the transition layer between the standard convection zone and its overshooting extension,
$z$ is the distance from this transition layer, $H_{v}$ the velocity scale height of
the overshooting convective elements at $z=0$, and $H_p$ the pressure scale height at the same point.
To establish the transition layer, first the Schwarzschild criterium is used to determine the edge of the convection 
zone and then the $z=0$ point is shifted a radius $0.01H_p$ (or $\fov H_p/2$ if $\fov<0.02$) into the convection 
zone where $D_0$ is computed. From $z=0$ and towards the radiative zone, the overshooting formulation is used. 
For convective cores lower than $0.001$ stellar masses overshooting is never added.
The same $\fov$ value was used for the core and the envelope and throughout the evolutionary sequence.

Different grids of models were used for each star. For KIC$\,$8006161 (and the related proxy) the grid is composed of evolution 
sequences with masses $M$ from $0.94M_{\odot}$ to $1.06M_{\odot}$ with a step of $\Delta M = 0.005M_{\odot}$, 
initial abundances [M/H] from  $-0.12$ to  $0.40$ with a step of $0.05$ and
mixing length parameters $\alpha$ from $1.5$ to $2.2$ with a step of $\Delta \alpha=0.1$. 
Since these stars are expected to have convective core only at the beging of the main sequence,
and adding overshooting can cause longer lived convective cores, no overshooting was considered in this case.

The initial metallicity $Z$ and helium abundance $Y$ were derived from [M/H], constrained by taking a Galactic chemical 
evolution model with $\Delta Y/\Delta Z= (Y_{\odot} - Y_0) / Z_{\odot}$ fixed. Assuming a primordial helium 
abundance of $Y_0=0.249$ 
and initial solar values of $Y_{\odot}=0.2744$ and $Z_{\odot}=0.0191$ (consistent with the opacities and GS98 abundances 
considered above) a value of $\Delta Y/\Delta Z= 1.33$ is obtained.
A surface solar metallicity of $(Z/X)_{\odot}=0.0229$ was used to derive values of $Z$ and $Y$ from the [M/H] interval. 

For a typical evolutionary sequence in the initial grid, 
we save about 100 models from the ZAMS to the TAMS. Owing to the very rapid change in the dynamical time scale of the models, 
$\tdyn=(R^3/GM)^{1/2}$, such grid is too coarse in the
time steps. Nevertheless, the dimensionless frequencies of the p-modes change so 
slowly that interpolations between models introduce errors much lower than the observational ones.
This procedure was discussed in more detail in \cite{2016A&A...591A..99P} and was found safe and consumes relatively less time.

For KIC$\,$9139163 the procedure is similar except that the grid is composed of masses from $1.32M_{\odot}$ to $1.60M_{\odot}$ 
with a step of $\Delta M = 0.01M_{\odot}$ and initial abundances [M/H] from  $-0.10$ to  $0.40$ with a step of $0.05\,$.
In this case overshooting was included, the parameter $\fov$ goes from $0$ to $0.04$ with a step of $0.01$.

A $\chi^2$ minimization, including p-mode frequencies and spectroscopic data, was applied to the grid of models. 
The general procedure is similar to that described in \cite{2016A&A...591A..99P}, but for these stars
all the eigenfrequencies are  approximately in the p-mode asymptotic regime and hence a simplified procedure can be done. 
Specifically we minimize the function
\begin{equation}
\chi^2=\frac{1}{3} \left( \chifreq+ \chidyn + \chispec \right)
\; .
\label{eq_chi2}
\end{equation}

Regarding the spectroscopic parameters, we have included the effective temperature $\teff$ the surface 
gravity $\log g$, the surface metallicity $Z/X$, and the luminosity $L$. 
Thus
\begin{equation}
\chispec= \frac{1}{4} \left[ \left( \frac{\delta \teff}{\sigma_{\teff}} \right)^2+ 
\left(\frac{ \delta (Z/X)}{\sigma_{ZX}} \right)^2+ 
\left(\frac{\delta g}{\sigma_{g}}\right)^2 + \left(\frac{L}{\sigma_{L}}\right)^2
\right]
\; ,
\end{equation}
where $\delta \teff$, $\delta (Z/X)$, $\delta g$ and $\delta L$ correspond to differences between the observations and the 
models whereas $\sigma_{\teff}$, $\sigma_{ZX}$, $\sigma_{g}$ and $\sigma_{L}$ are their respective observational errors.
Their values are given in Table~\ref{table:1}. 

The term $\chifreq$ in Eq.~\eqref{eq_chi2} corresponds to the frequency differences between the models and the observations 
after removing a smooth function
of frequency in order to filter out surface effects not considered in the modelling. 
This surface term is computed only using radial oscillations, and after scaling the frequency differences with the dimensionless energy $I_{nl}$, namely, 
\begin{equation}
I_{n0} \frac{\delta \omega_{n0}}{\omega_{n0}} \to  S (\omega) = B_0 + \sum_{i=1}^k B_i P_i(x)
\; ,
\label{eq_surfb}
\end{equation}
where $B_i$ are constant coefficients, $P_i$ a Legendre polynomial of order $i$, and $x$ corresponds to 
$\omega$ linearly scaled to the interval $[-1,1]$. A value of $k=2$ was adopted for the same reason than 
in \cite{2016A&A...591A..99P}.

Whence the surface term is determined, we consider radial as well as non-randial modes for computing the corresponding 
minimization function
\begin{equation}
\chifreq = \frac{1}{N-k-1} \sum_{j=1}^N \left( \frac{ \delta\omega_j/\omega_j - I_j^{-1} S (\omega_j) }
{\sigma_{{\omega}_j}} \right)^2 
\; ,
\label{eq_minfreq}
\end{equation}
where $j=1,\ldots N$ runs for all the modes and 
$\sigma_{{\omega}_j}$ is the relative error in the frequency $\omega_j$.

The term $I_j^{-1} S (\omega_j)$ subtracted to the relative frequency differences in Eq.~\ref{eq_minfreq} includes a 
constant coefficient which contains information on the mean density of the star. In fact, introducing the dynamical 
time $\tdyn=(R^3/GM)^{1/2}$ one has
$\delta \omega_{n0}/\omega_{n0} = \delta \tdyn/\tdyn + \delta \sigma_{n0}/\sigma_{n0}$
where $\sigma_{nl} =\omega_{nl}/\tdyn$ are the dimensionless frequencies.
As in \cite{2016A&A...591A..99P} we restore it in the minimization procedure adding a new $\chi^2$ term, but increasing the error
compared to the formal one derived from individual frequencies since this term 
can also include some influence from surface effects as discussed in \cite{2016A&A...591A..99P}.
Specifically we introduce the quantity
\begin{equation}
\chidyn = \left( \frac{B_0 - B_{00}}{\sigma_{B_0}} \right)^2
\label{eq_chi2a0}
\; ,
\end{equation}
where $B_{00}$ is an offset fixed such that $\nu^\mathrm{model} -  \nu^{\mathrm{obs}}$ would be positive within errors for the 
lower radial modes.
The parameter $\sigma_{B_0}$ is the error associated to $B_0$ and was taken as $\sigma_{B_0}=B_{00}\,$. Typically
we found values $\sigma_{B_0}\sim 10^{-3}$, at least one order of magnitude higher than the formal error found in a 
typical fit to Eq.~\eqref{eq_surfb}. 

To estimate the uncertainty in the output parameters we assumed normally distributed uncertainties for the observed frequencies, 
for $\sigma_{B_0}$, and for the spectroscopic parameters. We then
search for the model with the minimum $\chi^2$ in every realization
and compute mean values and standard deviations.

\section{Results}

\subsection{Best-fit model without frequency shifts}

We computed the best-fit model as described above for the two real stars and for the proxy stars using as input the spectroscopic parameters of Table 1 and, for the real stars, the observed frequencies and, for the proxies, the theoretical frequencies modified according to the description of Section 2 respectively.
Table~\ref{table:2} summarizes the results for the two stars and their proxies for the case where no frequency changes due to
magnetic activity are added.
The first row corresponds to KIC$\,$8006161 and the second one to its proxy. As it can be seen, 
the results have slightly lower errors for the proxy than for the actual star. 
The third row corresponds also to the proxy but in this
case the frequency and spectroscopic errors were halved. As it can be seen in Table~2 the resulting uncertainties in the 
stellar parameters are reduced by the same order of magnitude.
Finally, for KIC$\,$8006161
we have also carried out a similar fit but considering only a subset of modes with $\ell < 3$ and frequencies in the
central range, with the lowest frequency errors, namely,
$\ell=0$ modes with $n=17-26$, $\ell=1$ modes with $n=15-27$ and $\ell=2$ modes with $n=15-22\,$ (the full frequency set was
given in the first paragraph of section~2).
The resulting model parameters are shown in the last row of Table~\ref{table:2}, 
and we can conclude that the estimated uncertainties are similar to those in the first row where the full set of frequencies 
were considered.

Rows 5, 6, and 7 in Table~\ref{table:2} correspond to KIC$\,$9139163, its proxy and its proxy with halved frequency errors. 
Here the results are similar in all the cases. 
The most relevant differences between the star and its proxy are the $\chi^2$ values, which are 
higher for the observations. This could be the result of the approximations and 
uncertainties of the physics considered in the evolution codes.
Some of these discrepancies between models and observations can come 
from the magnetic activity and in principle one could wonder if frequency shifts induced by the magnetic activity could be 
detectable through higher $\chi^2$ values compared to other stars of the same type. 

\subsection{Adding the frequency shifts}

\begin{figure}[h!]
\begin{center}
\includegraphics[scale=0.5]{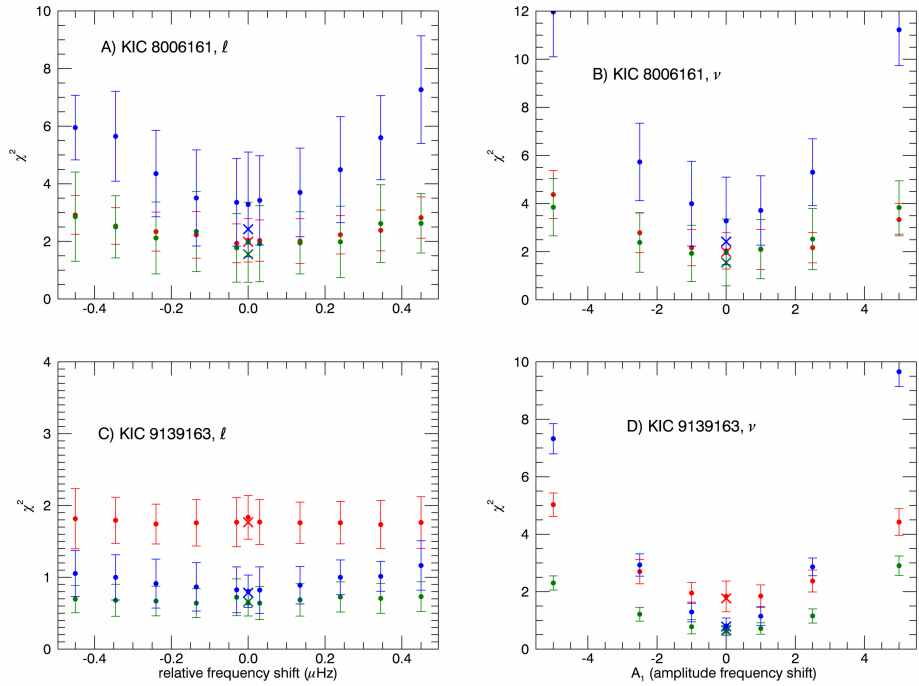}
\end{center}
\caption{Merit function $\chi^2$ as a function of the frequency shift introduced for KIC$\,$8006161 (upper panels) and for 
KIC$\,$9139163 (lowe panels). 
Here $\ell$ or $\nu$ indicates if the shift considered was $\ell$ or $\nu$ dependent. Red points 
correspond to adding the frequency shifts to the actual stellar data,
green points correspond to adding the frequency shifts to the proxy with the same frequency errors and blue points 
correspond to the proxy with half of the observed frequency errors. The magnitudes
shown in the $X$-axes are explained in Section 3.1 and 3.2 (see also Section 5.2).}
\label{fig:5}
\end{figure}

We then added the frequency shifts with the $\ell$-dependent frequency shift and with the frequency dependent shift. The merit functions, $\chi^2$, are shown in Fig.~\ref{fig:5} where panel A (resp. C) corresponds to KIC$\,$8006161 (resp. KIC$\,$913916) with the shift that depends on $\ell$ and panel B (resp. D) corresponds to KIC$\,$8006161 (resp. KIC$\,$913916) where the shift varies with frequency.
On the one hand, for the $\ell$-dependent cases the $X$-axis corresponds to the difference in $\mu$Hz between the radial and non 
radial modes. For instance a value of $0.3\,\mu$Hz corresponds to a shift of $\delta\nu =0.2\,\mu$Hz for the radial oscillations 
and $\delta \nu =-0.1\,\mu$Hz for non-radial oscillations. In addition a constant shift was also included as indicated
in section~3.2. The circles with error-bars at $x=0$ correspond to shifts that only include the constant term whereas the 
crosses at $x=0$ correspond to the results without frequency shifts.

On the other hand for the $\nu$-dependent cases the $X$-axis is the 
amplitude $A_1$ of the oscillatory function introduced in Eq.~\ref{Eq:1}, as factors of the actual values found for each 
star (see Figure~\ref{fig:4} for a graphic representation of the frequency shifts corresponding to $x$-values of $1$ in panels
B and D of Figure~\ref{fig:5}). At $x=0$ there are again two values per star: the circle which include the $A_0$ term and 
the cross corresponding to the result without frequency shifts.

Let us consider the simulations in panel A. The red points correspond to the $\chi^2$ values obtained by adding the frequency
shifts to the actual stellar data and errors. In this case $\chi^2$ increases with the 
absolute value of the frequency shift introduced but the changes are  within errors and could hardly indicate an incorrect 
fit even for the highest values (frequency shifts of $0.45\,\mu$Hz).
For the proxy star with the same frequency errors (green points) we obtain a similar behaviour.
However for the proxy star with half the frequency errors (blue points) 
the $\chi^2$ values clearly increase even when only the constant shift is considered; for
frequency shifts higher than $0.3\,\mu$Hz they are above the 1-$\sigma$ level. These kind of frequency shifts is within the 
range of observed values as shown in Section~3, but presumably they will be only present 
at a given time of the magnetic activity.

The results for the frequency-dependent shifts (panel B of Figure 5) show that for the extreme changes (that is an oscillatory frequency shift with an amplitude 
5 times those observed in KIC$\,$8006161) 
even when considering the  actual stellar parameters and errors (red points) the $\chi^2$ increases significantly. 
For the proxy with half the errors (blue points) even amplitudes of $2.5A_1$ could lead to an incorrect model fit.

For KIC$\,$9139163 the results on the oscillatory dependent term are similar (panel D) to the first star, 
however for the $\ell$-dependent case (panel C) it seems that an acceptable fit is always obtained. 

As a conclusion, an analysis based on the merit function $\chi^2$ cannot identify a bad model fit if we introduce 
a $\nu$-dependent shift with an amplitude $A_1$ for the oscillatory component smaller that 2.5 times the 
observed one. On the other hand, any value 
considered in this work for the $\ell$-dependent shift can be identified as a bad fit, except for the $\sim 1M_{\odot}$ case 
(KIC$\,8006161$) and assuming frequencies errors half the observed ones. 

\subsection{$\ell$-dependent frequency shifts}

Despite the different $\chi^2$ values found above, the stellar parameters derived for the proxy with the full errors and half 
errors gives very similar results, so from now on we will only show results for the proxy with half the actual errors. 

\begin{figure}[h!]
\begin{center}
\includegraphics[scale=0.45]{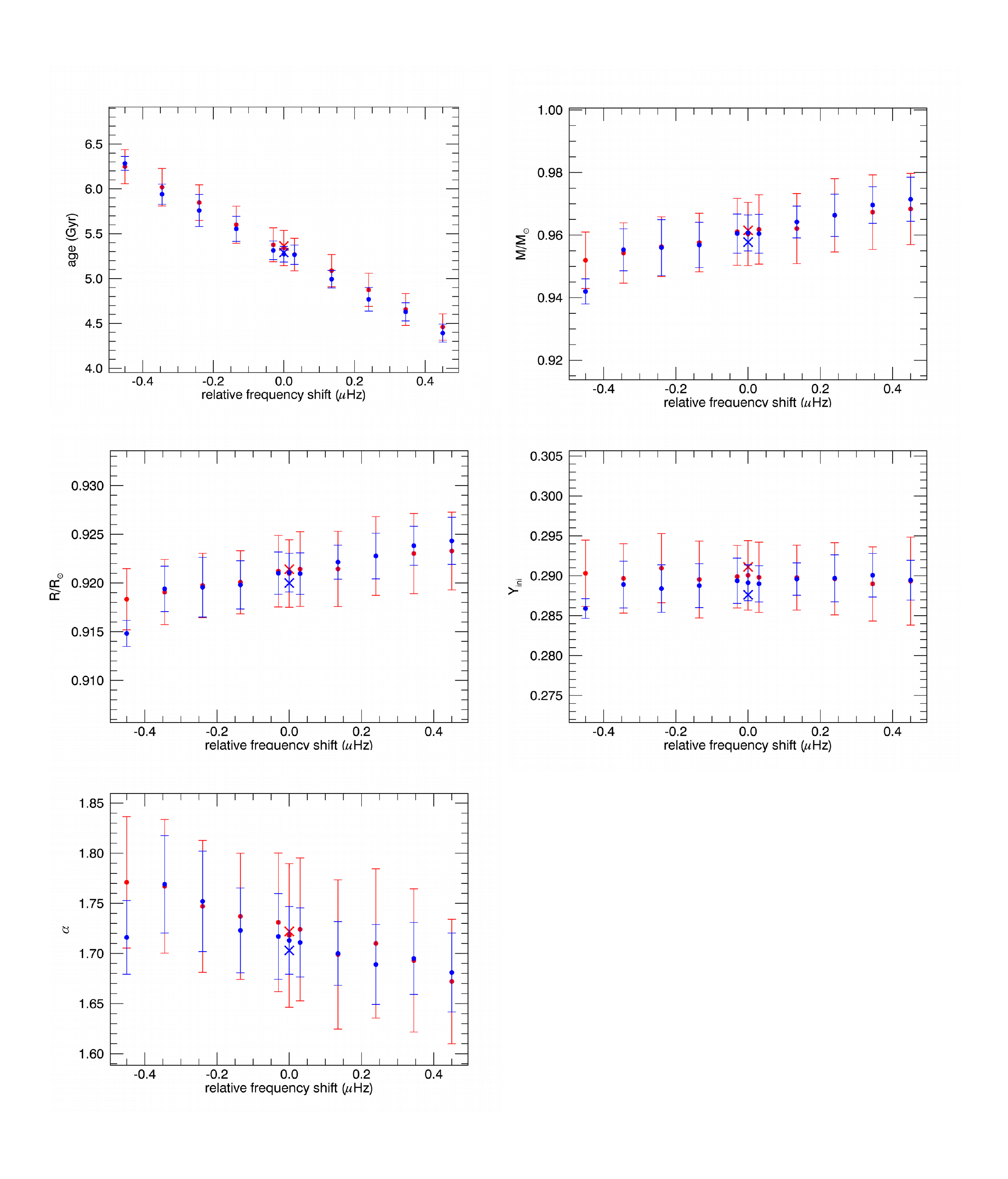}
\end{center}
\caption{Stellar parameters derived from the minimization procedure as a function of the $\ell$-dependent frequency shift introduced in the frequencies. Red points are for KIC$\,$8006161 whereas blue points corresponds to the proxy.}
\label{fig:6}
\end{figure}

We first derived the stellar parameters from the minimization procedure after introducing the $\ell$-dependent frequency shift 
(see Figure~\ref{fig:6} for KIC$\,$8006161).
For some of the parameters the changes induced by the frequency shifts 
are higher than the formal uncertainties. In particular for the age we find a clear trend with a decrease of $\sim 3.5\%$ 
every $0.1\,\mu$Hz of increase in the difference between the frequency shifts of radial and non-radial modes. 
Thus, for such a star where different $\ell$ are experiencing different frequency shifts due to magnetic activity, 
the age estimate can be more than 10\% away from the real age of the star. 
This can be compared with the uncertainties of $3\%$ when using the the actual frequencies and errors as well as 
the $1\%$ uncertainty for the proxy with half frequency errors.
Qualitatively, these results can be understood
in terms of the associated change in the small separation that when using spherically symmetric models can only be interpreted 
in terms of changes in the stellar core.
For the mass and radius the changes are smaller, of the same size than typical (formal) uncertainties. 
Specifically there is an increase of $\sim 1\%$ per $0.1\,\mu$Hz in the mass and of $\sim 0.3\%$ per $0.1\,\mu$Hz in the radius
that are of the same order as the formal uncertainties given in Table~2. 

\begin{figure}[h!]
\begin{center}
\includegraphics[scale=0.45]{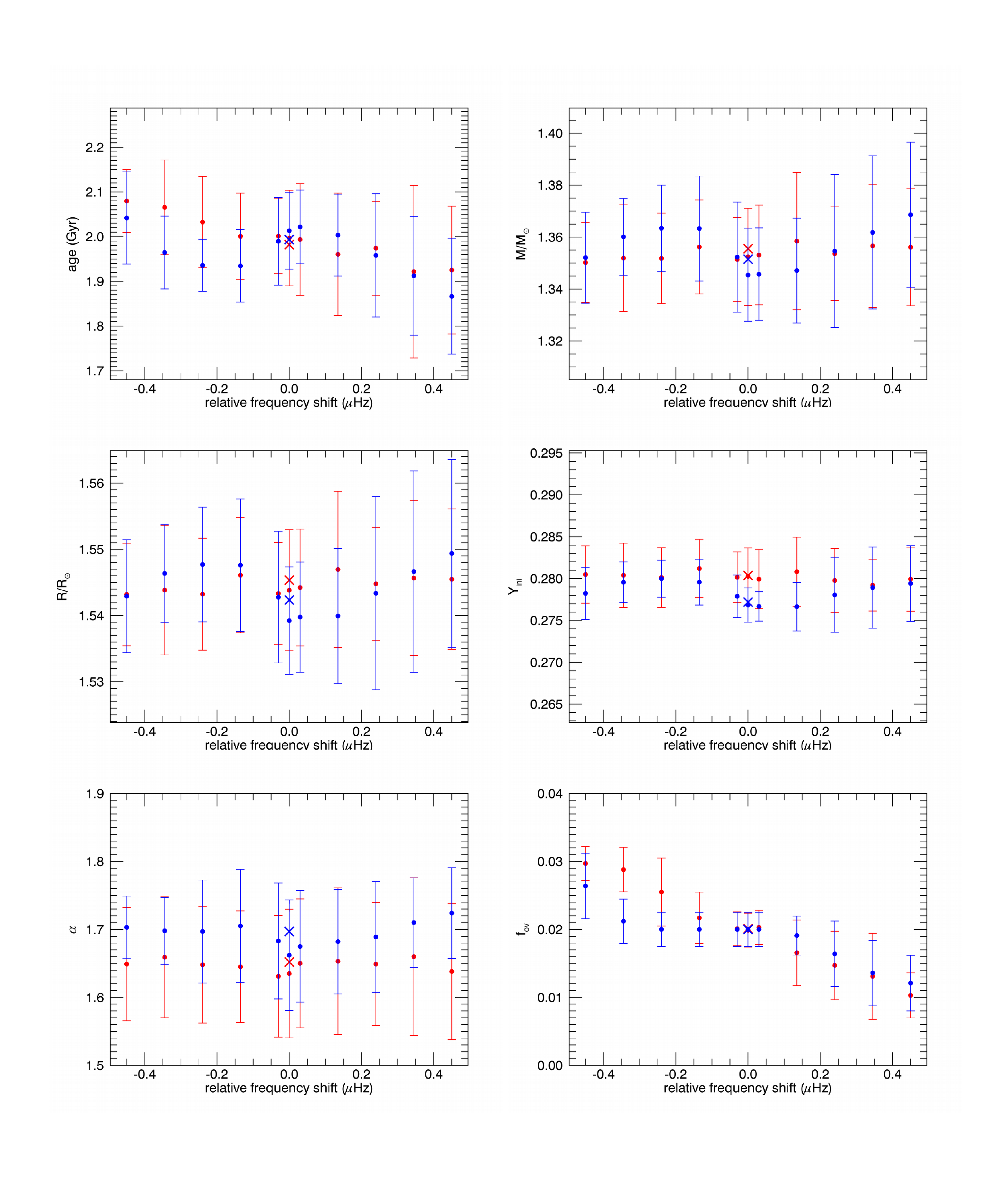}
\end{center}
\caption{Stellar parameters derived from the minimization procedure as a function of the $\ell$-dependent frequency shift introduced in the frequencies. Red points are for KIC$\,$9139163 whereas blue points corresponds to the proxy.}
\label{fig:7}
\end{figure}

Fig.~\ref{fig:7} shows the results when an $\ell$-dependent frequency shift is introduced in the eigenfrequencies of 
KIC$\,$9139163. In this case, the frequency shifts do not change the results by 
more than the formal uncertainties, except perhaps for the overshooting parameter $\fov$ for which we find changes of 0.02, above the 1-$\sigma$ level.

\subsection{frequency dependent shifts}

\begin{figure}[h!]
\begin{center}
\includegraphics[scale=0.5]{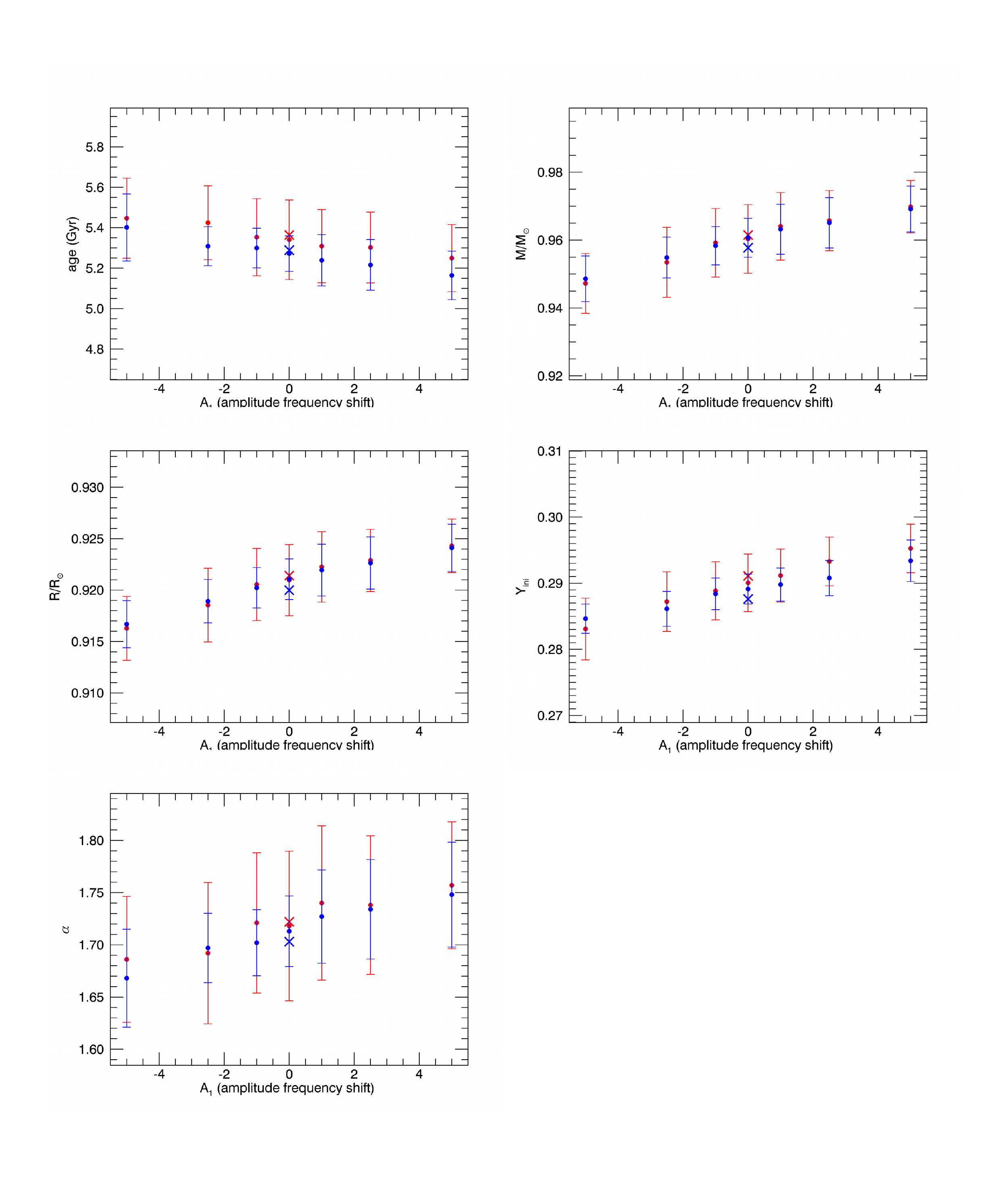}
\end{center}
\caption{Stellar parameters derived from the minimization procedure as a function of the $\nu$-dependent frequency shift introduced in the frequencies. Red points are for KIC$\,$8006161 whereas blue points corresponds to the 
proxy.}
\label{fig:8}
\end{figure}

Let us now consider the frequency dependent changes discussed in Section~3. In Fig.\ref{fig:8} we show the results for 
KIC$\,$8006161. Remember that in this case $X$-axis correspond to the amplitude $A_1$ in Eq.~\ref{Eq:1} in times the 
$A_1$ value derived from the observations. There are two points with $A_1=0$, the cross corresponding to the modes 
without any 
frequency shift ($A_0=0$ in Eq.~\ref{Eq:1}) and the dot corresponding to a frequency shift with the constant coefficient 
$A_0\ne 0 $ derived by \cite{2018A&A...611A..84S} but with the oscillatory term removed.

In this case, the changes in the age estimate are within the error bars.
However, there is a clear increasing trend in the mass, radius, 
and initial helium abundance. Specifically by taking $A_1$ equals to the observed value, the age decreases by $0.4\%$,
the mass increases by $\sim 0.6\%$,
the radius by $\sim 0.15\%$, and the helium abundance by $\Delta \yini \sim 0.002$ ($1\%$). 
These numbers are in practice below typical errors and could hardly be discriminated from other source of errors. 
In principle for large oscillatory amplitudes (higher values of $A_1$) the changes could be more relevant, 
$\Delta \yini \sim 0.01$ ($5\%$). 
However, as discussed early, these larger changes would result in high $\chi^2$ values, which would allow us to identify an 
incorrect model fit. 

\begin{figure}[!htb]
\begin{center}
\includegraphics[scale=0.5]{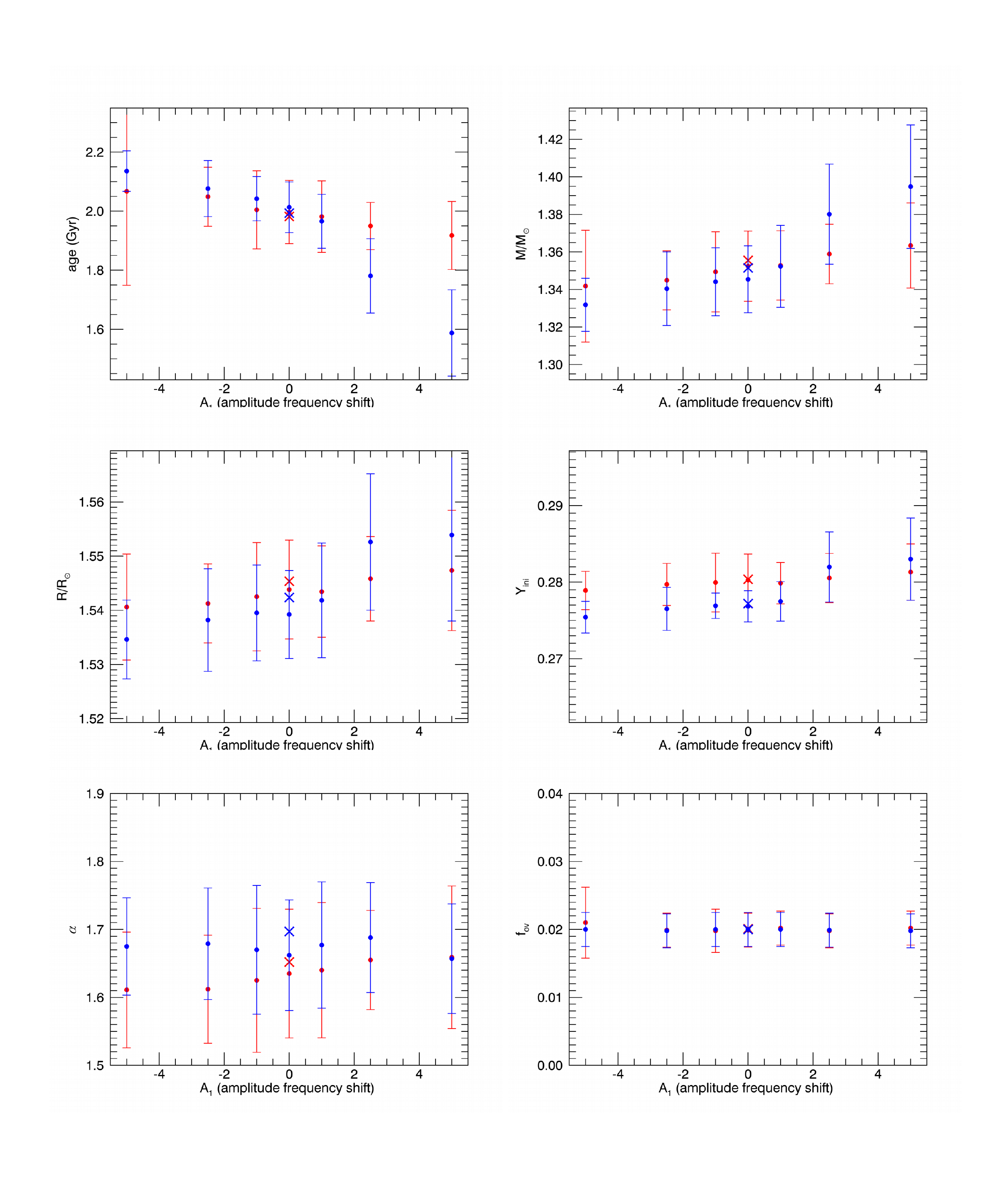}
\end{center}
\caption{Stellar parameter derived from the minimization procedure as a function of the $\nu$-dependent frequency shift introduced in the frequencies. Red points are for KIC$\,$9139163 whereas blue points corresponds to the 
proxy.}
\label{fig:9}
\end{figure}

Figure~\ref{fig:9} shows the results for KIC$\,$9139163. In this case the frequency shifts introduced do not 
change the results by more than the formal uncertainties though like KIC~8006161, similar trends in the age, mass, and radius 
can be seen.

\section{Conclusions}

Quantifying the influence of the frequency shifts caused by the magnetic activity on the determination of the stellar parameters
with asteroseismic techniques cannot be expressed with a simple mathematical rule because most of the techniques rely on some 
kind of model fitting that includes different constraints on a multiple parameter problem. 
For instance, quite different results could be obtained if one deals with the overshooting by introducing
a free parameter or if it is fixed a priori. The same can be true when considering the helium and metallicity abundances, they 
can be implemented as two independent parameters or they can be coupled through a $\Delta Y / \Delta Z$ enrichment law. 
Moreover, if rather than a direct comparison between observed and theoretical frequencies, derived quantities are used
to determine some specific parameters (e.g. see \cite{2019MNRAS.483.4678V} for using the acoustic glitches to determine the
helium abundance) then the sensibility to the magnetic perturbations can also be different.
In this work we have not tried to consider a broad range of situations but rather limit the analysis to some few 
examples, which we believe can be representative of some model fitting procedures. Although generalizing our quantitative results
to other cases should not be correct, we hope that our results can at least serve as order-of-magnitude guidance for a large 
range of cases.

If observations are limited to periods of months, magnetic activity can cause mode frequencies to depart from their raw 
values by tenths of $\mu$Hz or even more. 
We found that an $\ell$-dependent shift due to magnetic activity similar to that reported in \cite{2018ApJS..237...17S}
can introduce errors in the age of the order of $10\%$ at most, and of a few percents in the mass and radius. 
These figures can be higher than formal uncertainties and in particular 
we found that the uncertainties were more relevant for  the $1M_{\odot}$ example 
than for the $1.4M_{\odot}$ case. 
In principle one can think that this result could be due to the different quality of the input parameters in particular because
KIC$\,$8006161 has a larger range of frequencies, including some $\ell=3$ modes, but we have checked that limiting 
the frequency modes of this star to a range similar to the observed one for KIC$\,$9139163 leads to very similar results.
Hence it seems that the different positions in the HR-diagram should be the cause of the higher influence of the magnetic activity on the determination of the stellar parameters (relative to errors).

As long as surface uncertainties remain significant in the models, the frequency dependence of the 
frequencies shifts induced by the magnetic activity would be partially masked by the filtering procedures required in the 
modelling. However as shown by \cite{2018A&A...611A..84S}, the frequency shifts can include an
oscillatory component of a period similar to that corresponding to the acoustic depth of the He~II zone, which would not be suppressed by the surface filtering.
These shifts can also give rise to a misdetermination of 
some global stellar parameters, as for instance the helium abundance. But we have found that the uncertainties introduced
seem to be below the $3\%$ level in all the parameter analyzed.

Reducing the uncertainties on the frequencies to half of the observed one improves the determination 
of the stellar parameters in one of the cases (KIC$\,$8006161) and potentially would allow to flag incorrect determination of 
the best-fit model as they seem to have high $\chi^2$ values. 


\section*{Funding}
This research was supported in part by the Spanish MINECO  under project ESP2015-65712-C5-4-R. The paper made use of the IAC Supercomputing facility HTCondor (http://research.cs.wisc.edu/htcondor/), partly financed by 
the MINECO with FEDER funds, code IACA13-3E-2493. R.A.G. acknowledge the support from PLATO and GOLF CNES grants. S.M. acknowledge support by the National Aeronautics and Space Administration under Grant NNX15AF13G, by the National Science Foundation grant
435 AST-1411685 and the Ramon y Cajal fellowship number RYC-2015-17697. ARGS acknowledges the support from National Aeronautics and Space Administration under Grant NNX17AF27G.


\bibliographystyle{frontiersinHLTH&FPHY} 
\bibliography{testmag}


\section*{Tables}

\begin{table*}[t]
\caption{Input stellar parameters used for the modeling of each star and its proxy. The frequency of maximum power $\numax$ and 
the large separation $\Delta \nu$ are in $\mu$Hz. Individual frequencies are taken from \cite{2012A&A...543A..54A}.}
\centering
\begin{tabular}{l c c c c c c}
\hline

star & $\numax$ & $\Delta \nu$ & $\teff$ & $\log g$ & [M/H] & $\log L/L_{\odot}$ \\ 

\hline
KIC$\,$8006161  & $3444$ & $149.9$ & $5325\pm 100$ & $4.34\pm 0.3\phantom{0}$ & $0.28\pm 0.2$ & $0.64\pm 0.04$ \\
PRO$\,$8006161  &        &         & $5349\pm \phantom{0}50$  & $4.49\pm 0.15$ & $0.40\pm 0.1$ & $0.62\pm 0.02$  \\
KIC$\,$9139163  & $1730$ & $\phantom{0}81.2$ & $6400\pm \phantom{0}84$ & $4.20\pm 0.2$ & $0.15\pm 0.09$ & $3.88\pm 0.69$  \\    
PRO$\,$9139163  &        &         & $6460\pm \phantom{0}42$ & $4.19 \pm 0.1$ & $0.255\pm 0.045$ &  $3.74 \pm 0.35$  \\

\hline
\end{tabular}
\label{table:1}
\end{table*}

\begin{table*}[t]
\caption{Stellar parameters derived for the modeled stars (KIC) and their proxies (PRO).} 
\centering
\begin{tabular}{l c c c c c c}
\hline
KIC & $\chi^2$ & $R$ $(R_{\odot})$ & $M$ $(M_{\odot})$ & Age (Gyr) & $Z/X$ & $\tau_{\rm{HeII}}$ $(s)$ \\
\hline

KIC$\,$8006161           & $2.0\pm 0.6$ & $0.921\pm 0.003$ & $0.961\pm 0.010$ & $5.36\pm 0.18$ & $0.047\pm 0.005$ & $597\pm\phantom{0} 2$ \\
PRO$\,$8006161 $\sigma$ & $1.6\pm 1.1$ & $0.920\pm 0.002$ & $0.957\pm 0.006$ & $5.27\pm 0.16$ & $0.041\pm 0.004$ & $598\pm\phantom{0} 2$ \\
PRO$\,$8006161 $\sigma/2$ & $2.4\pm 1.3$ & $0.920\pm 0.001$ & $0.958\pm 0.003$ & $5.29\pm 0.06$ & $0.043\pm 0.003$ & $598\pm\phantom{0} 2$ \\
KIC$\,$8006161 0--2    & $2.1\pm0.6$  & $0.920\pm 0.003$ & $0.957\pm 0.009$ & $5.42\pm 0.17$ & $0.045\pm 0.005$ & $598\pm\phantom{0} 2$ \\
\\
KIC$\,$9139163        & $1.8\pm 0.3$ & $1.545\pm 0.010$ & $1.356\pm 0.020$ & $1.98\pm 0.12$ & $0.034\pm 0.004$ & $823 \pm 12$\\
PRO$\,$9139163 $\sigma$ & $0.7\pm 0.2$ & $1.544\pm 0.010$ & $1.355\pm 0.021$ & $1.97\pm 0.10$ & $0.032\pm 0.003$ & $819 \pm 14$\\
PRO$\,$9139163 $\sigma/2$ & $0.8\pm 0.2$ & $1.542\pm 0.009$ & $1.352\pm 0.020$ & $1.99\pm 0.09$ & $0.030\pm 0.003$ & $818 \pm \phantom{0}9$\\
\hline
\hline
\end{tabular}
\label{table:2}
\end{table*}

\section*{Figure captions}

\end{document}